\documentclass[runningheads]{llncs}
\usepackage[utf8]{inputenc}

\usepackage{hyperref}
\usepackage{algorithm}
\usepackage{algpseudocode}
\usepackage{pgffor}
\usepackage{listings}

\usepackage{tikz}
\usetikzlibrary{shapes,arrows}

\usepackage{comment}
\usepackage{amsmath}
\usepackage{amssymb}
\usepackage{bsymb}

\usepackage{oz}
\usepackage{cleveref}
\usepackage{subcaption}
\usepackage{xcolor}
\usepackage{todonotes}

\usepackage{wrapfig}

\definecolor{orcidlogocol}{HTML}{A6CE39}
\usetikzlibrary{svg.path}
\tikzset{
  orcidlogo/.pic={
    \fill[orcidlogocol] svg{M256,128c0,70.7-57.3,128-128,128C57.3,256,0,198.7,0,128C0,57.3,57.3,0,128,0C198.7,0,256,57.3,256,128z};
    \fill[white] svg{M86.3,186.2H70.9V79.1h15.4v48.4V186.2z}
                 svg{M108.9,79.1h41.6c39.6,0,57,28.3,57,53.6c0,27.5-21.5,53.6-56.8,53.6h-41.8V79.1z M124.3,172.4h24.5c34.9,0,42.9-26.5,42.9-39.7c0-21.5-13.7-39.7-43.7-39.7h-23.7V172.4z}
                 svg{M88.7,56.8c0,5.5-4.5,10.1-10.1,10.1c-5.6,0-10.1-4.6-10.1-10.1c0-5.6,4.5-10.1,10.1-10.1C84.2,46.7,88.7,51.3,88.7,56.8z};
  }
}
\renewcommand{\orcidID}[1]{%
  \resizebox{8px}{8px}{
      \href{https://orcid.org/#1}{\tikz[yscale=-1,transform shape]{\pic{orcidlogo}}}}%
}

\begin{document}

\title{Trace Refinement in B and Event-B\thanks{This research presented in this paper has been conducted within the IVOIRE project, which is funded by ``Deutsche Forschungsgemeinschaft'' (DFG) and the Austrian Science Fund (FWF) grant \# I 4744-N. The work of Austrian authors has been partly funded by the LIT Secure and Correct Systems Lab sponsored by the province of Upper Austria.}}
%
%
\author{Sebastian Stock\inst{1}\orcidID{0000-0002-2231-8656} \and
Atif Mashkoor\inst{1}\orcidID{0000-0003-1210-5953} \and
Michael Leuschel\inst{2}\orcidID{0000-0002-4595-1518} \and
Alexander Egyed \inst{1}\orcidID{0000-0003-3128-5427}
}
\authorrunning{S. Stock et al.}
%
\institute{Johannes Kepler University, Linz, Austria\\
\email{firstname.lastname@jku.at} \and 
Heinrich Heine University, Düsseldorf, Germany\\
\email{lastname@hhu.de}
}
\maketitle              
\begin{abstract}
Traces are used to show whether a model complies with the intended behavior.
A modeler can use trace checking to ensure the preservation of the model behavior during the refinement process.
In this paper, we present a trace refinement technique and tool called \textit{BERT} that allows designers to ensure the behavioral integrity of high-level traces at the concrete level.
The proposed technique is evaluated within the context of the B and Event-B methods on industrial-strength case studies from the automotive domain.
\end{abstract}

\section{Introduction}
\label{sec:introduction}
In a correct-by-construction model development process,  refinement takes center stage. The idea is to incrementally enrich a model in detail (horizontal or vertical refinement) while preserving the correctness of already introduced functionalities and properties. This correctness assurance process requires both verification and validation. While proof obligations (POs) \cite{abrial2005b} can ensure that the model is verifiable, validation obligations (VOs) \cite{mashkoor2021validation} can be used to assess the validity of the model. 

A trace can ascertain two conditions: 1) a specific end-state is reachable after executing steps in a specific order, and 2) a specific order of transitions is feasible.
When creating a trace for a model, the designer may be interested in keeping the specific trace throughout the refinement chain as evidence that particular behavior is preserved.
However, the time invested in creating the trace may be lost during the refinement process.
Furthermore, the trace may no longer be replayable due to a change in the operations or the addition of new operations.
One could argue that the designer could postpone validation until the model is concrete or recreate the trace on the refined machine. However, the first approach is against the spirit of early validation in the modeling process,
 while the latter case entails additional work and complexity. 

In this paper, we present a technique and tool called \textit{BERT} (\underline{B} and \underline{E}vent-B Trace \underline{R}efinement \underline{T}echnique) showing how traces on abstract models can be refined automatically to replay on later stages of the refinement chain.
With this technique and tool, we aim to make traces transferable between an abstract and refining machine in the context of B \cite{abrial2005b}, and Event-B \cite{abrial2010modeling}.
The ability to transfer traces increases the worth of traces in multiple aspects.
For one, we can assure ourselves of the presence of abstract behavior in the refinement, which helps us validate the specification. Additionally, we encourage tracing and the exploration of the model via animation in general.
With BERT, designers can automatically transfer the findings of their experiments to the next stage of development, i.e., the next refinement step. This is an aspect of early validation, which is desirable when writing formal models. 
However, we aim not to establish a refinement relationship but to transform a trace from an abstract machine $M_A$ to a concrete machine $M_C$.
So trace refinement in our context means not \textit{establishing a refinement relationship via traces} but \textit{refinement of a trace}. Trace refinement enables {\it comfortable} early validation of models via trace checking. We choose B and Event-B as candidate formal methods in this endeavor. The proposed technique is implemented as an add-on for the animator and model checker ProB \cite{leuschel2008prob}. The proposed technique is validated using industrial-strength case studies from the automotive domain.

The rest of the paper is structured as follows. \Cref{sec:background} introduces formal methods B and Event-B and the tool-set ProB. \Cref{sec:soundness} discusses the relationship between refinement checking and trace refinement and formalizes the technique of this paper. \Cref{sec:BERT}  introduces the techniques used for trace refinement. In \Cref{examples_and_evaluation} BERT is evaluated on two mid-size Event-B case studies. In \Cref{lessons_learned} the observations regarding algorithm and case studies are presented. Section \ref{related_work} compares the presented technique with the related work. Finally, Section~\ref{future_work} concludes the paper.

\section{Formal methods B and Event-B}
\label{sec:background}
The B modeling language, and its successor Event-B, shown in \Cref{fig:abstract}, are formal methods based on set theory and first-order predicate logic.
Although the methods differ in various ways, as explained by Leuschel~\cite{leuschel2021spot} and Mashkoor et al.~\cite{mashkoor18a}, both promote the correct-by-construction development paradigm. Designers can describe a machine that behaves like a state automaton with both languages.
In B, so-called \texttt{operations} are used to make transitions between states, while in Event-B, the transitions are made via \texttt{events}. Both events and operations manipulate the state of the machine\footnote{In the following, events and operations will be used as interchangeable words.}. The events have guards consisting of predicates checked against the machine's state. Depending on whether a guard is met, the transition is enabled. With the events, we can manipulate variables written in the \texttt{variables} section of the machine. For a machine, we can also define \texttt{invariants} that constrain the state space and are also used to define variables.

\subsection{Refinement}
\label{subsec:refinement}
In B and Event-B, refinement is a way to enrich the model by adding new variables or events or by replacing existing variables with more complex constructs. The key to refinement is that the concrete model preserves abstract behavior. Refinement enhances the abilities of a model without violating its properties.

In classical B, refinement of operations happens in a strict 1:1 relation. Practically this means that every operation from the abstract machine is refined in the concrete one only once, and no new operations are introduced. Furthermore, the name of the operation and the number and name of its parameters cannot change in the concrete machine $M_{C}$ once defined in the abstract machine $M_{A}$. 

In Event-B, refinement is more liberal.
For example, it is allowed to rename events, and it is allowed to introduce new events in $M_{C}$ that refine the invisible \texttt{skip} event from $M_{A}$.
Furthermore, unlike B, in Event-B, multiple events in $M_{C}$ can refine one abstract event from $M_{A}$, and the parameters can also be changed.
To check for the consistency of an Event-B refinement, one can use proofs showing a refinement does not violate the abstract model. These proofs are supported by, for example, the Rodin platform~\cite{abrial2010rodin}. 

\begin{figure}
     \centering
     \begin{subfigure}{0.4\textwidth}
         \centering
         \includegraphics[width=\textwidth]{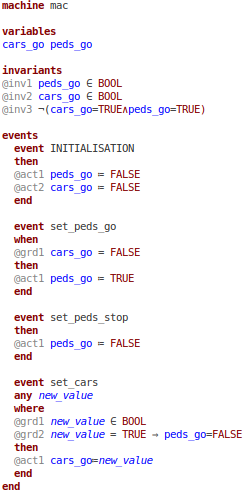}
         \caption{Abstract machine with binary light status}
         \label{fig:abstract}
     \end{subfigure}
     \hfill
     \begin{subfigure}{0.5\textwidth}
         \centering
         \includegraphics[width=\textwidth]{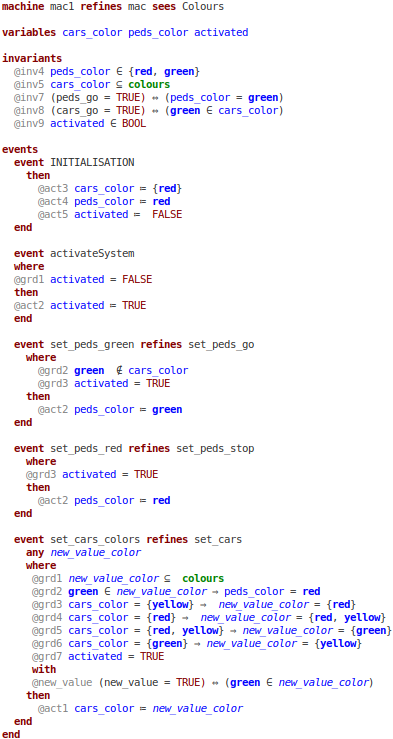}
         \caption{Refined machine with colors}
        \label{fig:concrete}
    \end{subfigure}
         \bigskip
     \begin{subfigure}{0.4\textwidth}
         \centering
         \includegraphics[width=\textwidth]{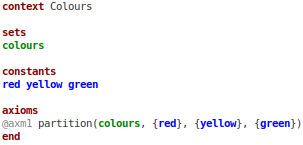}
         \caption{Context for colors}
         \label{fig:context}
     \end{subfigure}\\
      \begin{subfigure}{\textwidth}
         \centering
        \includegraphics[width=0.9\textwidth]{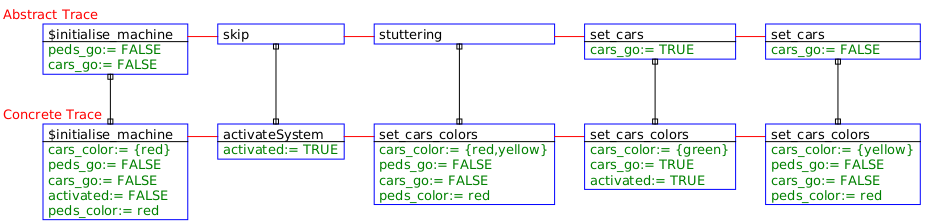}
\caption{Example of trace refinement result visualization outputted by BERT}
\label{fig:traceRefinement}
     \end{subfigure}

     \caption{Traffic light refinement example}
\label{fig:traffic}
\end{figure}

\Cref{fig:abstract} shows an abstract model of a traffic light taken from the Rodin handbook~\cite{EventBHandbook}. We see two variables, one for pedestrians and one for cars, stating whether they can pass. We can manipulate these variables with events. Additionally, the invariant \textit{@inv3} ensures that cars and people cannot pass at the same time. 

\Cref{fig:concrete} shows the refinement of \Cref{fig:abstract}. Here, two main features are implemented. First, we replace the Boolean variables with actual colors introduced in \Cref{fig:context} and couple the abstract and concrete states with \texttt{@inv7} and \texttt{@inv8}, thus ensuring that the safety invariant established in \Cref{fig:abstract} is also present in \Cref{fig:concrete}.
Furthermore, we have two colors for pedestrians and four for a car traffic light. The fact that these four phases\footnote{We model them in the way how German traffic lights operate.} have a specific order is accounted for in \texttt{set\_cars\_colors} where we have four consecutive phases for the lights. namely: \texttt{green} $\rightarrow$ \texttt{yellow} $\rightarrow$ \texttt{red} $\rightarrow$ \texttt{red,yellow} $\rightarrow$ \texttt{green}. Second, we introduce a new behavior in the form of the \texttt{activateSystem} event that refines \texttt{skip}. This event symbolizes the start of the system and is fired once. 

In Event-B, stuttering events are allowed. Stuttering describes the phenomenon when one abstract transition/state is replaced with multiple concrete ones. From the perspective of the abstract model, the concrete one has the same behavior. From the concrete model perspective, one may have to do multiple steps in the concrete state-space to mirror one abstract step.
The abstract handling of the traffic light contains two consecutive phases for cars \texttt{TRUE} $\rightarrow$ \texttt{FALSE} and the concrete version contains four consecutive color phases, as mentioned earlier. Abstract and concrete states are coupled via the witness \texttt{@new\_value}, such that the light is \texttt{green} whenever the Boolean variable would be \texttt{TRUE} and vice versa, thus mapping one abstract state to multiple concrete ones.

\subsection{ProB}
ProB~\cite{prob2Ui} is a model checker and animator for formal methods, including B and Event-B. ProB allows for sophisticated manipulation and reasoning about models of different formal languages with its extensions and libraries, making it a suitable tool for trace refinement. BERT is implemented as an add-on to ProB and utilizes its infrastructure, e.g., Java APIs of the kernel, to execute the necessary tasks.

\section{Trace refinement concept}
\label{sec:soundness}
Back and Wright~\cite{back1994trace} showed how to establish a refinement relationship by showing that the behavior of the concrete machine simulates the abstract machine. 
We are not establishing a refinement relationship here but transforming a trace from $M_A$ to $M_C$. The target of trace refinement is to find a trace in $M_C$ that contains the behavior of the trace from $M_A$. This is achieved step-wise by transforming the abstract trace's transitions and adapting them for the concrete machine. However, as Back and Wright's approach works well for transforming a concrete trace back to an abstract one, one cannot simply conclude the other way around. This is because not every abstract trace has a concrete counterpart. Similarly, only because an abstract trace can be refined for a concrete machine, those machines are not necessarily in a correct refinement relationship. However, the concrete machine mimics the behavior of the abstract machine encoded in the form of a trace.

\subsection{Trace refinement in B}
\label{subsec:descriptionB}
For B, the concept of trace refinement is implemented in the form of refinement checking to ensure that an abstract machine contains all concrete traces~\cite{butler:zum97,dunne,leuschel2005automatic}. We define a trace $T$ as an ordered list of transitions $t_1,\ldots,t_n$, which describe the operation's name, the parameters used, and the state reached. For a B machine, it is refined to $T'$ iff every transition $t$, consisting of a state and an operation, from the abstract trace has 1) a distinguishable counterpart in the corresponding concrete trace, 2) transitions and states have the same order, and 3) the length of the abstract trace and the concrete trace is equal. These attributes are directly derived from the notion of refinement of B refinement and the findings of Leuschel and Butler~\cite{leuschel2005automatic} regarding refinement checking via traces.

\subsection{Trace refinement in Event-B}
\label{sec:descriptionEventB}

Trace refinement in Event-B is commonly established via proofs. 
In terms of our trace refinement algorithm, it is more challenging than the B version. 
 Let us consider a small example trace for \Cref{fig:abstract}. \texttt{INITIALISATION} $\rightarrow$ \texttt{set\_cars(TRUE)} $\rightarrow$ \texttt{set\_cars(FALSE)} represents the abstract model allowing the cars to go and stops them afterwards. Now we want to refine the trace and adapt it to \Cref{fig:concrete}. Consider \Cref{fig:traceRefinement} as a reference to a successful refinement\footnote{The graphical output shown in \Cref{fig:traceRefinement} is the result of BERT's automatic visualization. As ProBs internal representation is used, some names, for example \texttt{$initalize\_machine$}, are held in the ProB style.}. \Cref{fig:traceRefinement} shows two traces, both to be read horizontally. Each rectangle represents a trace transition, with the upper part being the operation and the body representing the changed values. The different transitions are connected via the red lines. The connection between the abstract and concrete transitions is marked with strokes with a square head on both ends, e.g., the concrete initialization contains a more sophisticated set of variables than the abstract one. Two placeholders \texttt{skip} and stuttering were inserted in the places where the concrete trace diverges from the abstract - those two transitions were not part of the abstract trace. The event \texttt{activateSystem} that refines \texttt{skip} has become a necessity in order to progress further in the trace. Same goes for \texttt{set\_cars\_colors}, which is a stuttering action. Note the values of the machine here. While it was possible to directly switch from \texttt{TRUE} to \texttt{FALSE} in the abstract version, a stuttering action is needed as \texttt{FALSE} is mapped to \texttt{red} and \texttt{red,yellow} in the concrete version. 

For a successful trace refinement, we borrow some of the attributes from~\cref{subsec:descriptionB}: 1) pairs of abstract and concrete states are distinguishable, and 2) order persevering. However, with 3), the trace might get lengthier as new observable behaviors might become necessary to mirror the abstract behavior in the concrete model. Though acceptable, being unavoidable, the addition of new behaviors is not preferable.

A significant challenge for the trace refinement algorithm is stuttering, skip refinements, and their ability to introduce divergence~\cite{derrick2018refinement}. Divergence means that the model neither runs in a deadlock nor terminates but evolves in a way not observable from an abstract standpoint. For refinement in Event-B, this can be the case as \texttt{skip} and \texttt{refines} keywords add behaviors not observable from an abstract viewpoint. We cannot be sure whether these behaviors terminate or deadlock at some point. Event-B offers a solution for this by annotating events as \texttt{convergent} associated with a proof that the event will occur only finite times often. Further challenges are that multiple events can refine a single one, thus forcing us to investigate all concrete events, and that renaming can occur.

To sum up: Trace refinement in Event-B, like in B, ensures that the abstract behavior is preserved in the concrete trace in the same order. However, there can be additional, new behaviors only observable in the concrete machine at any point in the concrete trace. If the related events are not annotated as \texttt{convergent} at some point, these concrete events can occur infinitely often. If the convergence is unproven, we can run into a situation where we often discover new concrete states that do not present in the abstract model, thus running into an endless loop without any abort condition.

\section{BERT}
\label{sec:BERT}
BERT\footnote{BERT is available as part of ProB2, see\\ \url{https://www3.hhu.de/stups/downloads/prob2/snapshot/}} is a technique and tool for trace refinement. It needs $M_A$, $M_C$, and the trace to be refined. BERT uses the ProB libraries to prepare the machines on the implementation level. The actual refinement process is done with the help of the ProB animator. The solution can be stored in the trace format used by the animator\footnote{A demo is provided at \url{https://doi.org/10.6084/m9.figshare.16909006.v2}}. It can be replayed on the concrete machine without additional measures. If there is no solution, an error message is displayed, and the last attempt is given as a trace. The visual representation provided by BERT is shown in \Cref{fig:traceRefinement}.

\subsection{Trace refinement for B}
\label{subsec:classicalB}
The existence of data refinement makes it necessary to observe abstract and concrete state space simultaneously when doing trace refinement as a designer can replace abstract variables with concrete ones. The consequence is that the abstract variables are no longer visible in the concrete machine. This could leave us unsure which transition to take when doing trace refinement, as each transition in the trace is formulated as a predicate over state variables and enabled operations. Missing information could lead to ambiguity. The problem is that machines are independent components. Therefore, we combine the state spaces of $M_A$ and $M_C$ to eliminate the data refinement problem. One way to achieve this is by merging abstract syntax trees (ASTs) and creating an intermediate machine. After a successful combination of state spaces, we can proceed with trace refinement.

Let $M(R,V,I,Op)$ be a simplified machine. Each of the four elements represents one of the ASTs node types. 
\begin{itemize}
    \item $R$ is the set of references to other machines, represented as an identifier list
    \item $V$ is the set of variables, represented as an identifier list
    \item $I$ is the invariant, a conjunction of predicates
    \item $Op$ is the set of operations. An operation has the form $op(N,In,Ou,B)$
        \begin{itemize}
            \item $N$ represents the name of the operation
            \item $In$ represents input variables of the operation
            \item $Ou$ represents output variables of the operation
            \item $B$ represents the body of the operation
        \end{itemize}
\end{itemize}
Let $M_A(R_A,V_A,I_A,Op_A)$ be the abstract and $M_C(R_C,V_C,I_C,Op_C)$ be the concrete machine. An interleaved machine $M_I = M_A || M_C$ has the form $M_I(R_A \cup R_C, I_A \land I_C, V_A \cup V_C, Op_A || Op_B)$ and the operations are interleaved by $Op_A || Op_B = \{op(N, In, Ou, B_1 || B_2) | op(N, In, Ou, B_1)\in Op_A \land op(N, In, Ou, B_2) \in Op_C\}$. In the case of the operation bodies, $||$ is the parallel substitution. As parameters cannot be changed, they stay untouched. In B machines with multiple, overlapping references, merging $R_A$ and $R_C$ can be challenging. This is because of the possibility that $R_A$ and $R_C$ could contain conflicting statements which would need to be resolved by renaming large parts of the AST to avoid conflicts. Currently, these cases are not treated in this work and should be considered as out of scope.

Algorithm \ref{alg:algB} shows the process of the actual trace refinement. For reasoning, we borrow the $start \overset{op}{\underset{M_C}\rightarrow} s$ notation from the work of Leuschel et al.~\cite{leuschel2005automatic} to describe the process of creating a transition. This notation expresses multiple things. First, a transition goes from a state $start$ to a state $s$. Furthermore, the transition uses the operation $op$. Finally, the transition is executed on the machine $M_C$. Internally, the transition is written as a predicate consisting of abstract variables, operation parameters, and the operation name. If an operation with the name exists that fulfills the predicates, the operation is seen as a valid transition.

\Cref{alg:algB} aims to implement the condition for successful trace refinement we set up in \Cref{subsec:descriptionB}: an order-preserving and length preserving image from an abstract transition to the concrete counterpart. In \Cref{alg:algB}, we show how B traces are refined. We take a transition, find its counterpart, and append the found concrete transition to our solution in each step. This happens in line 12 of \Cref{alg:algB}. As multiple solutions per transition are possible, we have to carry all those solutions with us, which result in the fact that we store all concrete traces that are candidates for being a solution in \texttt{currentTraces}. Due to the way we copy \texttt{resultTraces} into \texttt{currentTraces}, we dismiss those traces from further calculations that are no more qualified candidates. The way we use the coupled state spaces, the $start \overset{op}{\underset{M_C}\rightarrow} s$ relation and reduce the trace in each step, we ensure that \Cref{alg:algB} produces the correct result.

\begin{algorithm}[t!]
\caption{Algorithm of BERT for B}
\label{alg:algB}
\begin{algorithmic}[1]

\Require $M_{C}$ that has access to $M_{A}$, Trace T made out of a list of transitions $t_1,\ldots, t_n$ with t = (op,s)
\State $resultTraces \leftarrow$ \Call{FindPath}{$\langle \langle  (\_,root)  \rangle \rangle$, T, $M_{C}$}
\If{resultTraces = $\langle \rangle$}
\State println(\emph{"Trace could not be translated"})
\Else
\State selectAnySolution(resultTraces) \Comment{Hand selection of the result}
\EndIf

\Function{FindPath}{currentTraces, T, $M_{C}$}

\For{(op, s) $\in$ T }

\State resultTraces $\leftarrow \langle \rangle$ 

\For{trace $\in$ currentTraces}
\State (\_, currentState) $\leftarrow$ last(trace)
\State trans $\leftarrow \{(op, s') | currentState \overset{op}{\underset{M_C}\rightarrow} s' \land s \sqsubseteq s' \}$  \Comment{Next reachable states}
\For{tran $\in$ trans} \Comment{Collect all solutions}
    \State resultTraces $\leftarrow$ append(resultTraces, append(trace, tran)) 
\EndFor
\EndFor
\State currentTraces $\leftarrow resultTraces$
\EndFor

\State \textbf{return} currentTraces
\EndFunction
\end{algorithmic}
\end{algorithm}

\subsection{Trace refinement for Event-B}
\label{subsec:eventB}
Merging state spaces for Event-B is analogous to B, as shown in~\cref{subsec:classicalB}, with minor changes in how parameters and 1:n relations between concrete events are handled. 

The trace refinement algorithm for Event-B is shown in \Cref{alg:algEB}. Besides static information about alternatives and skip refinement, it includes a list of \texttt{trace} data structures, which itself is a triple and consists of the current concrete trace, the abstract transitions left, and the already visited nodes of the state space. We loop until we either have no further state to traverse and return unsuccessful or find a trace that refines all abstract steps and has found a solution. Inside the loop, we consider each \texttt{trace} structure, which is currently maintained as a possible solution for all valid successors. In line 13 of~\Cref{alg:algEB}, one can recognize the formulation already used in~\Cref{alg:algB} where a direct equivalent to an abstract transition exists. In this case, the abstract trace is reduced in size in line 16 of~\Cref{alg:algEB}, and the solution for this step is added to the set of solutions. In line 18 of~\Cref{alg:algEB}, we create all solutions for this step, erasing all stored states as we made progress in reducing the task size. New is the calculation for \texttt{skip} and stuttering in line 20. We remove every transition we have already seen from the found solutions to avoid being livelocked as we made no progress in reducing the task size. In line 23, we then construct the set of solution candidates. All solution candidates are then sent to the next iteration. As~\Cref{alg:algEB} terminates when it reaches the first set of solutions, we can be sure that we find a minimal solution without added behavior if there is any.

\begin{algorithm}[ht!]
\caption{Algorithm of BERT for Event-B}
\label{alg:algEB}
\begin{algorithmic}[1]

\Require $M_{C}$ that has access to $M_{A}$, Trace T made out of a list of transitions $t_1,\ldots, t_n$ with t = (op,s), a map \textit{alt} mapping abstract to concrete operations, the list of transitions refining \texttt{skip}
\State $resultTrace \leftarrow$ \Call{FindPath}{$\langle trace( \langle(\_,root) \rangle, T, seenTrans) \rangle ,  M_{C}, alt, skip$}
\If{resultTraces = $\langle \rangle$}
\State println(\emph{"Trace could not be translated"})
\Else
\State return selectAnySolution(resultTraces)
\EndIf

\Function{FindPath}{currentTraces, $M_{C}$, alt, skip}
\While{ $\lvert currentTraces \rvert \neq 0 \land \nexists trace \in currentTraces \centerdot \lvert trace(1) \rvert = 0$ }

\State resultTraces $\leftarrow \langle \rangle$

\For{entry $\in$ currentTraces}
\State (op, s) $\leftarrow$ first(entry(1))
\State (\_, cS) $\leftarrow$ last(entry(0))
\State t $\leftarrow \{(op', s') | op' \in alt(op) \land cS \overset{op'}{\underset{M_C}\rightarrow} s'\ \land s \sqsubseteq s'\}$
\For{tran $\in$ t}
 \State T $\leftarrow$ entry(1)
 \State newT $\leftarrow$ drop(T,0) \Comment{Reduce the task size}
 \State newTrace $\leftarrow$ append(entry(0), tran)
 \State resultTraces $\leftarrow$ append(resultTraces, trace(newTrace, newT, $\langle \rangle$)
\EndFor
\State t $\leftarrow \{(op',s') | (op' \in$ skip $ \lor op' \in alt(tOp)) \land 
cS \overset{op}{\underset{M_C}\rightarrow} s' \land s' \sqsubseteq cS\}$
\State t $\leftarrow$ t $\cap$ entry(2) \Comment{Remove all solutions leading into livelock }
\For{tran $\in$ t}
 \State seenTrans $\leftarrow$ tran $\cup$ entry(2) \Comment{Add seen transitions }
 \State newTrace $\leftarrow$ append(entry(0), tran)
 \State mewEntry $\leftarrow$ trace(newTrace, entry(1), seenTrans)
 \State resultTraces $\leftarrow$ append(resultTraces, newEntry)
\EndFor
\EndFor
\State currentTraces $\leftarrow$ resultTraces

\EndWhile

\State \textbf{return} $\{trace | trace \in currentTraces \land \lvert trace(1) \rvert = 0\}$  
\EndFunction

\end{algorithmic}
\end{algorithm}

\subsection{Practical limitations and optimizations}
\label{subsec:limits}
Practically the state space explosion problem can hinder both algorithms from finding a solution if the machine running the algorithm runs out of memory. In the case of Event-B, introduced divergence can result in an endless loop. However, it is up to the designer to show the convergence of the events.

The designer may limit the search depth or the number of investigated branches to mitigate the state space problem. In our implementation and the resulting testing, we had good experiences with this. A further modification to the proposed algorithms was a mixed breadth/depth-first search resulting in quicker results in our test cases for the price that, in rare cases, not the shortest solution was found. Further optimization was comparing candidate sub-traces for their equality, resulting in pruning traces that occurred multiple times.

\section{Case studies}
\label{examples_and_evaluation}
In this section, we experiment to validate the proposed technique on two industrial-strength case studies, both implementing models of an automotive system~\cite{caseStudy}. For brevity, we demonstrate only the Event-B implementation as shown in~\Cref{alg:algEB} on case studies as Event-B refinements pose a more significant challenge than B as far as trace refinement is concerned. In the first case study, we work out the quality aspect of trace refinement. In the second case study, we address the performance aspect.

\subsection{Pitman arm controller}
\paragraph{Case study description.}
The first case study is the pitman arm controller model \cite{leuschel2020modelling}. The authors modeled a pitman arm that controls car lights and the user's interaction with it. The model consists of three machines forming a refinement chain of: 
\begin{equation*}
  BlinkLamps \rightarrow PitmanController \rightarrow PitmanControllerTime 
\end{equation*}
\textit{BlinkLights} models the light environment of a car and its behavior. It allows controlling lights. Furthermore, one can set a blinking timer emulating the actual number of blinks of the lights. In the model itself, the amount is set via parameters of the \texttt{SET\_REMAININGBlinks} and its refinements. -1 means forever, 3 means three-time, and so on. The third feature is the tip blinking or comfort blinking of modern cars, which is also modeled. The \textit{PitmanController} extends the \textit{BlinkLights} and adds an actual car environment building on top of the lamps. The position of the Pitman arm and car's key state is modeled. In \textit{PitmanControllerTime} the abstract time counting is replaced with a concrete one. Normally the tip blinking of a car has a specified time interval, usually three to five times. \textit{PitmanControllerTime} models this behavior. For the experiment we will create a trace on \textit{BlinkLamps} and refine it step-wise to \textit{PitmanControllerTime}. We will present all three traces and evaluate what the refinement did to the trace and which knowledge one can extract from this.

\begin{figure}
     \centering
     \begin{subfigure}{0.4\textwidth}
         \centering
         \includegraphics[width=\textwidth]{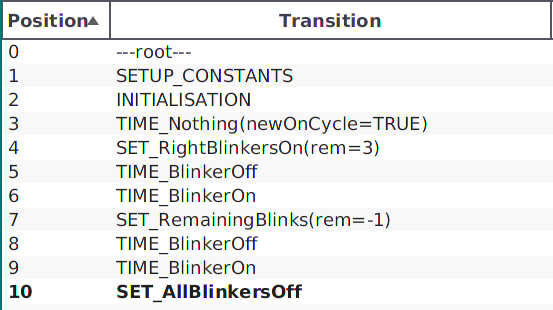}
         \caption{Original trace on BlinkLamps}
         \label{fig:blinkLamps}
     \end{subfigure}
     \hfill
     \begin{subfigure}{0.4\textwidth}
         \centering
         \includegraphics[width=\textwidth]{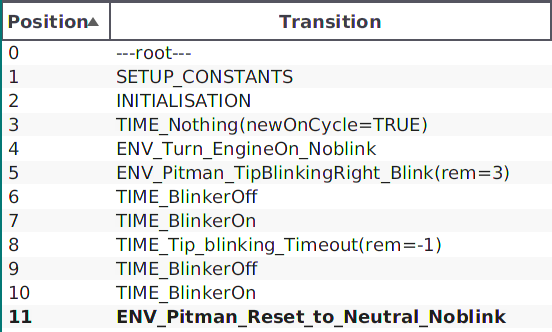}
         \caption{Refined trace on PitmanController }
         \label{fig:pitmanController}
     \end{subfigure}
         \bigskip
     \begin{subfigure}{0.4\textwidth}
         \centering
         \includegraphics[width=\textwidth]{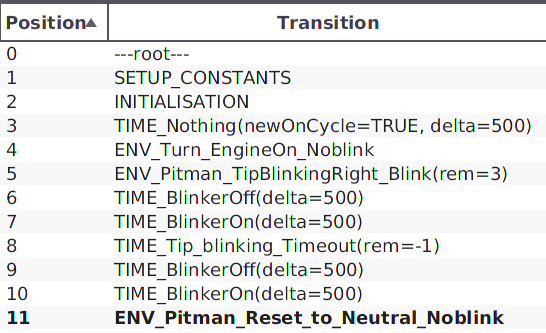}
         \caption{Refined trace on PitmanControllerTime}
         \label{fig:pitmanController2}
     \end{subfigure}
     \caption{Trace before and after the first and second step of trace refinement.}
     \label{fig:traceRefinementStudy}

\end{figure}

\paragraph{Results.}
In \Cref{fig:traceRefinementStudy} one can see the abstract trace and its refinements. The abstract trace in \Cref{fig:blinkLamps} captures the idea of first using the tip blinker, but the user then decides to make the blinking permanent after one blink. Again, after one blink, the user turns off the blinker entirely. 

In \Cref{fig:pitmanController} one can see the adaptation. The most important change is that it is now necessary to turn on the engine, for the lamps to blink. A new transition at position 4 is introduced for this. Renaming introduced with the refinement is considered, and the events have been changed accordingly, for example, \texttt{SET\_REMAININGBlinks} is now called \texttt{TIME\_Tip\_Blinking}, \texttt{SET\_RightBlinkersOn} is now called \texttt{ENV\_Pitman\_TipBlinkingRight\_Blink}, and \texttt{SET\_AllBlinkersOff} is now called \texttt{ENV\_Pitman\_Reset\_to\_Neutral\_Noblink}.

For the third refinement, the introduction of time had a minor effect on the existing behavior. This is mirrored in the resulting trace adaptation. In \Cref{fig:pitmanController2} we see no change compared to \Cref{fig:pitmanController} besides additional parameters for the events, for example, on position 3.

We gathered multiple insights into the effectiveness of trace refinement from our experiments. 1) The engine needs to be turned on for the lights to blink, which can be checked against the specification to learn if it is intentional behavior. 2) The newly introduced event in \textit{PitmanControllerTime} does not interfere with existing behavior. Once again, one can check if this is intentional. 3) From the perspective of animation, there is an event \texttt{SET\_RemainingBlinks} that can have parameters 1,2,3 as input. We observed that 2 always leads to a failing adaptation process. When checking for the reason, it was found that the state in which the remaining blinks are equal to 2 is no longer reachable. Therefore, all behaviors relying on this are no longer possible. Designers might want to check whether this is an intentional behavior with the stakeholders.

\subsection{Automotive adaptive exterior light system}
\paragraph{Case study description.}
The second case study is more complex than the first one and is about the specification of the adaptive exterior light system of an automobile~\cite{mammar2020event}. The case study focuses on the lights, their behavior, reaction to the user, and environmental changes. As a result, the refinement chain is more prolonged, reaching from \textit{M0} as the most abstract machine to \textit{M5} as the most concrete machine. 
\begin{itemize}
    \item \textit{M0} introduces a simple event with the name \textit{headLightSet} with the ability to set a specific light of the car to a specific level.
    \item \textit{M1} introduces the interfaces allowing the user to interact with the car.
    \item \textit{M2} introduces details to the direction indicators, the hazard lights, and the emergency brake light.
    \item \textit{M3} introduces details for the low beam lights.
    \item \textit{M4} introduces details for the cornering lights.
    \item \textit{M5} introduces details for the high beam lights.
\end{itemize}
The introduction order of the features in \textit{M2} up to \textit{M5} was arbitrary. 
This is related to the problem of expressing topological orders in the strict Event-B refinement hierarchy even tho there are attempts, such as by Mashkoor et al. \cite{mashkoor11b}. We provide an overview of the exact metrics of all refinement steps in~\cref{tab:refinementChainOther}.
Like the previous case study, the first refinement step sees a massive increase of events as several new features are introduced to model the user interface. The table shows that we have a continuous, stark growth of variables in each step, on average twelve, which heavily increased the state space in each step. The fact of the possible inputs further blows up the state-space for each event. As events had large sets of parameters, each step had a large set of possible parameter values. The resulting increase of the state space caused problems as generating all possible inputs for an event became very memory intensive. Thus, there was a need for cutting away solutions to not run out of memory. For this, the internal standard of ProB was used that provides the first five correct solutions to a predicate if not further modified. For the experiment, we create a longer trace (30 steps) and find an adaptation for each machine. As mentioned earlier, this quantitative experiment shows how well BERT deals with larger state spaces.

\begin{table}
\centering
\caption{Refinement chain statistics}
\label{tab:refinementChainOther}
\begin{tabular}{l|cccccc}
Machine              & M0  & M1 & M2 & \multicolumn{1}{l}{M3} & \multicolumn{1}{l}{M4} & \multicolumn{1}{l}{M5} \\ \hline
Introduced Variables & 1   & 15 & 9  & 12                     & 12                     & 12                     \\
Overall Variables    & 1   & 16 & 25 & 37                     & 49                     & 61                     \\
Introduced Events (via \texttt{skip})    & 1   & 1  & 0  & 0                      & 0                      & 3                      \\
Refined Events       & N/A & 10 & 3  & 5                      & 0                      & 0                      \\
Extended Events      & N/A & 1  & 11 & 14                     & 20                     & 20                     \\
Overall Events       & 1   & 12 & 14 & 19                     & 20                     & 23                    
\end{tabular}
\end{table}

\paragraph{Results.} With the introduced restrictions, a trace could be successfully adapted. The results are shown in \Cref{tab:resultsSecond}. For our experiment, we measure four categories: 1) the input trace to the algorithm, 2) the time needed for adaptation, 3) the number of transitions that need to be added for a successful adaptation, and 4) the amount of  solutions found for the input trace. 

The trace was not refined for M1 as it was created on this machine, thus N/A. We can see that the number of transitions stays constant while the calculation time increases up to M4. The reason here seems to be the number of possible solutions and the increasing number of variables later leading to state space explosion. The decline in time for M5 can be explained because there were not many solutions to be explored. No transitions were added, which was expected as the refinement steps focused on adding new behaviors, so the abstract behavior was preserved.

Overall we conclude the experiment a success. We anticipated problems with larger state spaces when designing the algorithm back in~\Cref{subsec:limits}. Nevertheless, we managed to refine a trace even under more challenging conditions.

\begin{table}[t!]
\centering
\caption{Trace refinement statistics for the trace refined with BERT}
\label{tab:resultsSecond}
\begin{tabular}{l|cccccc}
Machine                & M1 & M2  & M3   & M4   & M5   \\ \hline
Trace Input Size      & N/A & 96  & 96   & 96   & 96   \\
Adaptation Time (ms)  & N/A  & 1663 & 2289 & 8268  & 1938  \\
Added Transitions     & 96  & 0   & 0    & 0    & 0    \\
Individual solutions  & N/A  & 3   & 4    & 10   & 2   
\end{tabular}
\end{table}

\subsection{Threat to validity }
\label{subsec:threats_to_validity}

The application of~\Cref{alg:algEB} to case studies poses an internal threat. We used the readily available implementations. While the first case study was created to be animated, the second was not, meaning that creating traces was far more difficult for the second case study. We created meaningful traces for the first case study, and the results had meanings we could reason. The second case study was not created with animation in mind. Hence the traces created to test were not always meaningful. In the first case study, we observed quality features of the behavior of trace refinement, while in the second one, we could only observe quantitative behaviors.

\section{Lessons learned}
\label{lessons_learned}
\begin{itemize}
    \item Trace early. Following this long-standing advice can indeed save from making poor design choices. Trace refinement can then help to design elegant refinement steps.
    \item  Keep the refinement step simple and small, i.e., introduce one feature per refinement as also suggested by Mashkoor et al.~\cite{mashkoor2011guidelines}. This will simplify the refinement process and help understand the reasons for a failing trace refinement.
    \item Do not create unnecessary choice points. Checking choice points created by newly introduced events and non-deterministic choices can increase calculation time significantly. This was discovered when searching for a valid configuration that holds throughout the refinement chain.
    Whenever there was no first best choice but many second-best choices, the path branched heavily, leading to long-running times. This case can be handled by keeping the number of choice points introduced in each refinement step small.
\end{itemize}

Overall, BERT helped clarify whether a specific behavior in $M_A$ is still valid in $M_C$. This seems particularly interesting from the designer's perspective as it can validate the assumption that refinement ruled out a particular behavior. In addition, the high level of details that a trace provides can help a designer understand why a specific behavior is not feasible in the refinement, thus identifying modeling flaws. 

Especially interesting when using trace refinement is what can be learned from a failing adaptation. There are two reasons for a failing adaptation. Either there is no adaptation or the adaptation is not achievable within the set memory limits. If we expect a successful adaptation, a manual inspection of the resulting trace of the failed adaptation can uncover design flaws. It can hint at questionable design from the animation perspective, as shown in the second case study, where technical limitations made it hard to find solutions. One might want to rethink the model to keep it easily animatable and traceable. Also, with manual inspection of failing results, cases were encountered where events were not sound and forgot to reset a specific variable leading to a failing adaptation.

\section{Related work}
\label{related_work}
We compare our technique to the related work's existing trace and scenario refinement techniques. We want to stress the difference between traces and scenarios. Even though the term 'scenarios' is heavily overused, we can extract a baseline of scenarios being high-level representations of intended behavior. With high-level, we mean only selected pre-selected variables and transitions of interest are observed. Traces, on the other hand, are more low level. As traces need to be precise, they contain all available information about states and transitions.

\paragraph{Trace refinement.} One important inspiration for this paper is how refinement is defined and established in state-of-the-art CSP~\cite{hoare1978communicating} algorithms~\cite{gibson2014fdr3}. In CSP, one specification can refine another in multiple ways, i.e., trace- and failure refinement. The refinement relationship is established by showing that both specifications are equivalent regarding the traces they share or do not share. The notion of 'hiding' transitions was especially interesting for our technique. In Event-B, we encountered a similar situation when stuttering refinement occurred. Whenever a transition or variables only occur in the concrete specification, we 'hide' them to treat the situation like a case of B-trace refinement to check whether the trace is still valid in the concrete refinement. 

\paragraph{Scenario refinement.} Arcaini and Riccobene~\cite{arcaini2019automatic} describe how to refine ASM scenarios, which are used as acceptance tests. 
Using their proposed tool, scenarios are translated into an LTL formula. Then $M_{C}$ is loaded into the tool-related model checker, the LTL formula is negated, and the model checker will try to find a counter-example. This counter-example can be re-translated into a scenario. The authors define two mapping procedures from scenario to LTL formulas, one that is strict and will produce 1:1 traces and one that can produce 1:n traces. The latter will add new behaviors. Besides the obvious difference that the proposed technique deals with ASMs, the authors seem not to encounter divergences as part of the concrete model. The ASM technique completely rely on its existing framework using an existing scenario language and the LTL model checker. In contrast, in our technqiue, we can perform calculations only based on a state-space representation, and we have chosen the one provided by ProB. Therefore, we provide a common technique for trace refinement and a tool. The advantage of our technique is that whenever low level properties are of interest we are able to refine them.

Scenarios for Event-B were introduced, e.g., by Fischer and Dghyam~\cite{CucumberEventB} where Cucumber \cite{wynne2017cucumber} is used as a description language. However, compared to our work, the refinement abilities described in the paper are minimal. The created tests are not further refined but statically generated for each model version. Another technique to Event-B scenarios, which catered for refinement, was proposed by Malik et al. \cite{EventBScenarioRefinement} where for Event-B models, scenarios were encoded in CSP and then refined. However, this technique suffers from the need to encode scenarios in CSP, which loses exact information about states as CSP only supports reasoning about transitions. Furthermore, the problems tackled by our work are only partially tackled in this work. The idea of scenario refinement was then advanced by Snook et al.~\cite{DslScenarioRefinement} where it is investigated how scenarios can be refined or applied to more abstract machines. The authors also suggested a tool \cite{snoSec} for scenario generation for Rodin that is currently not able to refine those scenarios.

\paragraph{Dual animation.}Our technique focuses on going from an abstract to a concrete level while the technique of Hallerstede et al.~\cite{hallerstede2010refinement,hallerstede2013validation} proposes the opposite. In their technique, a concrete machine is linked to its predecessor, thus giving insight into how the abstraction behaves when the concrete model does something. This helps to explore the actual behavior of an abstract and concrete machine and the changes due to refinement. However, checking the preservation of behavior is a cumbersome, manual task and not a real option as too many details need to be considered. With our technique, this task is automatically done by the provided tool.

\section{Conclusion and future work}
\label{future_work}
This paper introduces BERT -- a technique and a tool to refine traces that are explicit paths through state space. With this tool, it is possible to refine traces of abstract B and Event-B machines for their respective refinement. This transfer helps designers to ensure the presence of an abstract behavior in the concrete specification. We successfully conducted experiments on two case studies to showcase the validity of the proposed technique. While BERT produced the expected results, the case studies helped find its role as a support tool for designers. With its help, one can comprehend if and why a specific behavior is still allowed in a refined model. The benefit for the designer is the fine-grained details a trace provides that allow exact reproduction of the defined behavior. 

We are currently tackling refinements for Event-B and using the usage of the \texttt{REFINES} keyword for B. In the future, we want to tackle B's alternative ways of enriching a machine's behavior, for example, \texttt{INCLUDES} or \texttt{EXTENDS}. This will leverage the burden of animating and refining B machines, making trace refinement through animation more convenient.

\bibliographystyle{splncs04}
\bibliography{bibliography}

\end{document}